\documentclass{iau} 
\usepackage{graphicx}

\title[S 344.~~The ISM and Star Formation in Dwarf Galaxies] 
{The Interstellar Medium and Star Formation in Dwarf Galaxies}

\author[Alberto D. Bolatto]   
{Alberto D. Bolatto$^1$}

\affiliation{$^1$Department of Astronomy, University of Maryland, \\ College Park, MD 20742, USA\\ email: {\tt bolatto@astro.umd.edu} \\}

\pubyear{2018}
\volume{344}  
\jname{Dwarf Galaxies: From the Deep Universe 
to the Present}
\editors{S. Stierwalt Editor \& K. McQuinn Editor, eds.}
\begin{document}

\maketitle

\begin{abstract}
This is a brief review of our understanding of the properties of the
    interstellar medium (ISM) in dwarf galaxies in connection to
    their star formation activity. What are the dominant phases of
    the ISM in these objects? How do the properties of these phases
    depend on the galaxy properties? What do we know about their cold
    gas content and its link to star formation activity? Does star
    formation proceed differently in these galaxies? How does star
    formation feedback operate in dwarf galaxies? The availability of
    observations from space-based facilities such as FUSE, Spitzer,
    Herschel, and Fermi, as well as observatories such as SOFIA and
    ALMA, is allowing us to make significant strides in our
    understanding of these questions. 
\end{abstract}

\firstsection 
\section{Introduction}

In this brief review I will focus on the mid- to small-scale properties of the interstellar medium (ISM) in dwarf galaxies. These scales are easiest to reach in two of our nearest neighbors, the Magellanic Clouds, located 52 and 63 kpc away. The typical corresponding scales for observations are then $\sim0.3$ parsecs per arcsecond, which afford uniquely detailed views of the ISM and star formation processes. By comparison, observations elsewhere within the Local Group will probe order-of-magnitude coarser scales, and observations probing extreme dwarf galaxies (such as I\,Zw\,18) outside the Local Group probe scales that are 300 times larger. Therefore much of my focus will be on Magellanic Cloud observations, particularly on the Small Magellanic Cloud (SMC), which is the smaller of the two and also the one with the lowest metallicity ($12+\log[{\rm O/H}]\sim8.0$, approximately $\sim0.2$ Solar). 

One of the recent revolutions in astronomy is the realization that all galaxies, including dwarf galaxies, are not closed systems. In fact throughout their cosmic histories these galaxies accrete material from their circum-galactic medium, merge with other galaxies, and expel enriched gas that lingers in their halos or escapes to the inter-galactic environment. If we were to follow a parcel of this gas through the process of star formation, we would find that it has to cool and recombine into neutral gas from its presumably mostly ionized state in the halo, become part of the cold phase for HI in the ISM, undergo a physical and chemical transformation where it is collected into denser regions that then become H$_2$ inside a molecular cloud, then experience processes inside the molecular cloud that further compress it to densities that enable it to efficiently collapse to successfully form a star. These processes all depend on the environmental conditions, and the type of galaxy where they occur.

\section{The Distribution of the Cold and Warm Neutral Gas Phases}

 In particular, forming the cold neutral medium is the required step prior to forming molecular gas. In the Milky Way, HI exists in cold and warm phases (CNM and WNM respectively) that appear to be in approximate pressure equilibrium in its midplane (pressure equilibrium is more debatable outside the plane where a larger fraction of the neutral hydrogen may be at pressures and densities outside equilibrium, \cite[Heiles \& Troland 2003]{Heiles:2003}; but see also \cite[Murray et al. 2018]{Murray:2018}). In the classical picture this distribution is set by thermal instability, brought about by the characteristic S-shape of the equilibrium cooling curve (\cite[Field, Goldsmith, \& Habing 1969]{Field:1969}; \cite[McKee \& Ostriker 1977]{McKee:1977}; \cite[Wolfire et al. 1995, 2003]{Wolfire:1995,Wolfire:2003}). 

This equilibrium cooling curve comes from considering the balance of all heating and cooling processes. In the Milky Way and for the CNM, these are dominated by the conversion of UV radiation into gas heating by the photoelectric effect in dust grains, and by the cooling primarily through [CII] $157.7$\,$\mu$m fine-structure emission. The interesting connection with dwarf galaxies is that ISM metallicity has a direct impact on the cooling and heating processes, possibly systematically altering the balance. Because of the mass-metallicity relation, heavy element abundances are strong functions of galaxy mass, and we systematically expect (and observe) dwarf galaxies to have much lower metallicities than larger systems (\cite[e.g., Tremonti et al. 2004]{Tremonti:2004}). 

Indeed, less abundant carbon directly translates into less efficient cooling per hydrogen nucleus. Because [CII] emission is such a dominant coolant over a wide range of densities and temperatures for the neutral medium it is not easily replaced, resulting in a rise in overall temperatures and a new equilibrium (\cite[Wolfire et al. 1995]{Wolfire:1995}). The heating processes, however, also experience changes. In particular the photoelectric heating (caused by the ejection of photo-electrons from the surfaces of dust grains impacted by  far-ultraviolet radiation, which collisionally thermalize with the surrounding gas sharing their energy and heating it), depends on dust-to-gas ratio, which is also a function of metallicity --- although likely in a non-linear way that may depend on the star formation history of the system (\cite[Draine et al. 2007]{Draine:2007}; \cite[Galametz et al. 2011]{Galametz:2011}; \cite[Remy-Ruyer et al. 2013]{Remy-Ruyer:2013}; \cite[Fisher et al. 2014]{Fisher:2014}). Moreover, the efficiency with which dust traps UV photons and ejects photo-electrons depends on properties of the dust grains themselves, such as their size distribution. There is strong evidence that these properties change with metallicity. For example, aromatic features attributed to polycyclic aromatic hydrocarbons (PAHs), which are the smallest dust grains, disappear from the mid-infrared galaxy spectra at low metallicities (\cite[e.g., Engelbracht et al. 2005, 2008]{Engelbracht:2005,Engelbracht:2008}). Whether this is caused by lower formation or enhanced destruction remains a matter of research (\cite[Gordon et al. 2008]{Gordon:2008}, \cite[Sandstrom et al. 2010]{Sandstrom:2010}), but interestingly other properties of these grains such as charge (which would affect their heating efficiency) appear to also change systematically in low metallicity environments (\cite[Sandstrom et al. 2012]{Sandstrom:2012}). 

Adding to this is the increasing evidence suggesting that small galaxies have a cosmic ray deficit. Although not important for the heating of gas that is exposed to UV, cosmic rays are likely to be the main source of heating and ionization in the dark, highly extincted central regions of molecular clouds where most star formation will take place (unless cosmic rays are heavily excluded from cloud centers). 
The fact that dwarf galaxies are likely to host a lower cosmic ray density was already suggested from the comparison of radio and far-infrared continuum luminosity measurements (\cite[e.g., Bell 2003]{Bell:2003}), but recent {\em Fermi} Large Area Telescope measurements have clearly shown that the Small Magellanic Cloud is under-luminous in gamma-rays, which are the result of the decay of pions produced by interactions between cosmic rays and interstellar gas (\cite[Abdo et al. 2010]{Abdo:2010}; \cite[L\'opez et al. 2018]{Lopez:2018}). These measurements imply an average cosmic ray density that is only 15\% that of the Milky Way, perhaps due to a smaller confinement volume or to removal by advection in winds that may have caused a substantial fraction of the cosmic rays generated to be lost (\cite[L\'opez et al. 2018]{Lopez:2018}). 

This adds up to a complex picture of the changing thermodynamics and chemistry in the neutral gas in dwarf galaxies. The effects on the balance of the warm and cold phases are, however, currently unclear. Emission-absorption measurements of the hyperfine transition of HI at 21~cm can be used to directly probe the temperature of the CNM and the fraction of cold gas in a system, although they are essentially limited to the Local Group (e.g., \cite[Heiles \& Troland 2003]{Heiles:2003}). Original measurements in the SMC (\cite[Dickey et al. 2000]{Dickey:2000}) found peculiar properties: the CNM is very scarce ($\lesssim15\%$ of the total HI), but its temperature is also very cold ($T\lesssim40$~K) with several lines of sight exhibiting temperatures of 20~K or colder, which would be more characteristic of Galactic cold molecular clouds. 

By comparison, in the Milky Way CNM accounts for about 40\% of all the HI, and its column-weighted mean temperature is closer to $T\sim100$~K, although clouds with $T\sim40$~K are not uncommon (\cite[Heiles \& Troland (2003)]{Heiles:2003}). These results for the SMC, however, were based on only 13 absorption detections, which highlights the importance of revisiting them as more powerful facilities become available. Preliminary results from a larger SMC HI survey with deeper integration and $\times3$ more detections suggest that some of these earlier results may have been caused by sensitivity biases, and the cold gas fraction is much closer to 30\% (Jameson et al., in prep.). The current results for the Large Magellanic Cloud (LMC) seem closer to the properties of the Galaxy, although they are also uncertain; the fraction of CNM may be similar to that of the Milky Way with a typical temperature perhaps somewhat colder (\cite[Marx-Zimmer et al. (2000)]{Marx-Zimmer:2000}).

\section{The Molecular Phase}

Basic photodissociation region theory shows that metallicity, mostly through the effect of the diminishing dust-to-gas ratio, has a strong effect on the structure of the molecular gas. Because H$_2$ is photodissociated by far-ultraviolet photons absorbed in the Werner and Lyman bands, which become optically thick even at fairly low column densities, the H$_2$ molecule is strongly self-shielding and its distribution is only a mild function of the extinction associated with dust (\cite[e.g., Wolfire et al. 2010]{Wolfire et al. (2010)}). The CO molecule, on the other hand, is not as heavily shielded and its survival relies on the extinction provided by dust, which decreases at least proportionally to decreasing metallicity (\cite[van Dishoeck \& Black 1988]{vanDishoeck:1988}). The result is that the column density required to turn most H into H$_2$ is increased only slowly by decreasing metallicity, while the column density required to turn C$^+$ into C and ultimately CO increases much more rapidly thus causing a rapid increase in the value of the CO-to-H$_2$ conversion factor (\cite[Wolfire et al. 2010]{Wolfire et al. (2010)}). As a consequence CO is a good tracer of H$_2$ at most metallicities in regions where the column density from the edge of the cloud results in an extinction that exceeds $A_V\sim1-2$, but when the metallicity drops these regions become increasingly rare, representing the column density peaks. These expectations are also reproduced by detailed computer simulations that include the H$_2$ and CO formation and photodissociation chemistry (\cite[e.g., Glover \& McLow 2011]{Glover:2011}, \cite[Glover \& Clark 2012a]{Glover:2012a}).

Note that measurements of the ionized carbon [CII] emission at $157.7$ $\mu$m strongly support a change in the CO-to-H$_2$ conversion factor at low metallicities that is consistent with theory and simulations. In particular, measurements in giant molecular clouds in the SMC are consistent with large, likely molecular envelopes surrounding CO-emitting cores. When [CII] is used to estimate the mass of H$_2$ in these envelopes (\cite[Jameson et al. 2018]{Jameson:2018}), the resulting ratio of CO emission to molecular mass versus extinction display trends very similar to the predictions from theory and simulations by \cite[Wolfire et al. (2010)]{Wolfire:2010} and \cite[Glover \& Clark (2012a)]{Glover:2012a}. Similar conclusions are reached by \cite[Pineda et al. (2017)]{Pineda:2017} modeling {\em Herschel} heterodyne [CII] measurements in the Magellanic Clouds. Any measurement of star formation per unit molecular gas mass in these systems needs to account for changes in the CO-to-H$_2$ conversion factor.

A consequence of these processes is that, at increasingly low metallicities, the HI shielding layer gets progressively thicker before H$_2$ can form. A measurable consequence is that the maximum surface density of HI observed in a galaxy must depend on its metallicity, if it reaches the column density necessary to form H$_2$. \cite[Schruba et al. (2018)]{Schruba:2018} perform this measurement in a collection of galaxies of different metallicities, finding that indeed the mean maximum HI surface density increases for decreasing metallicity. We can also test the behavior of CO. Indeed, because bright CO requires a minimum extinction, we would expect the relation between the brightness of CO emission and $A_V$ to be essentially independent of metallicity. \cite[Lee et al. (2015)]{Lee:2015} compare measurements of this relation in Perseus, the Pipe Nebula, and an ensemble of Milky Way molecular clouds to that in the SMC, finding that I$_{\rm CO}$ versus $A_V$ is essentially the same, with bright CO emission requiring $A_V\gtrsim1$. Both of these tests reinforce the idea that smaller, metal-poor galaxies, are increasingly less efficient at converting cold HI into H$_2$, but they are proportionally even more deficient in producing CO emission because of their lack of dust and associated shielding. 

Aside from the brightness of their CO emission and associated quantities, it has been difficult to find strong systematic differences between the resolved properties of the molecular clouds in these systems and those in more massive galaxies. In particular, the size-line width relation, which is defined on CO-emitting regions but is otherwise independent of the CO emission, appears reasonably consistent across many different systems ranging from WLM at the lowest metalicity (\cite[Rubio et al. 2015]{Rubio:2015}) to the Magellanic Clouds (\cite[Bolatto et al. 2008]{Bolatto:2008}; \cite[Hughes et al. 2013]{Hughes:2013}), NGC\,6822 (\cite[Schruba et al. 2017]{Schruba:2017}) and our own Milky Way. It is, however, clear that the individual molecular clouds identified in CO are systematically smaller and have lower surface brightness than in more metal-rich systems. 

\section{The Efficiency of Star Formation}

Against this backdrop, dwarf galaxies appear to be extremely efficient at turning molecular gas into stars, at least as judged on the basis of their CO emission (\cite[e.g., Schruba et al. 2012]{Schruba:2012}).
This is highlighted by the few systems that have been detected in CO at metallicities similar to or below that of the SMC: WLM (\cite[Rubio et al. 2015]{Rubio:2015}), DDO 70 (\cite[Shi et al. 2016]{Shi:2016}), and Kiso 5639 (\cite[Elmegreen et al. 2018]{Elmegreen:2018}) are prime recent examples where the ratio of star formation activity to molecular mass inferred from CO emission is disproportionately higher than that observed in more massive galaxies.  

The arguments outlined in the previous section, however, highlight the fact that CO is a poor tracer of the bulk of the H$_2$ in these systems. Is the apparent increase in efficiency (defined as star formation rate per unit molecular gas mass) purely explained by a change in the CO-to-H$_2$ conversion factor? If so, stars may be forming in gas that, while molecular, is at a low enough $A_V$ to not have any associated CO emission. Or are there changes in efficiency that are also happening in these systems?

There are a few such measurements of molecular star formation efficiency in the Magellanic Clouds. Using far-infrared dust emission measurements together with 21~cm HI data to estimate the amount of molecular gas, \cite[Bolatto et al. (2011)]{Bolatto:2011} and \cite[Jameson et al. (2016)]{Jameson:2016} find that the efficiency on large scales, using H$\alpha$ emission to estimate star formation activity, is similar to that observed in larger galaxies. Similar conclusions were reached by \cite[Schruba et al. (2017)]{Schruba:2017} in the Local Group dwarf NGC6822, with metallicity intermediate between both Magellanic Clouds.

It is interesting, however, to use a more localized star formation tracer for these studies, since the H$\alpha$ emission has a significant contribution from a diffuse component in the Magellanic Clouds that makes it specially difficult to interpret in on scales of hundreds of parsecs. \cite[Ochsendorf et al. (2017a)]{Ochsendorf:2017a} uses LMC observations of both massive young stellar objects (YSOs) identified in the mid-infrared, as well as H$\alpha$ emission, and compares them to gas masses derived from CO and from dust (the latter taken from the \cite[Jameson et al. 2016]{Jameson:2016} analysis). They study, among other things, the star formation efficiency per free-fall time, which is a standard measure of efficiency (\cite[e.g., Krumholz et al. 2012]{Krumholz:2012}). What they find is a trend with cloud mass that is very similar to that observed in the Milky Way by \cite[Lee et al. (2016)]{Lee:2016}, albeit with a very large scatter in both the Galactic and the LMC samples. 

An analysis using YSOs is being carried out in the SMC by Johnson et al. (in prep.), based on {\em Hubble} Space Telescope observations. This analysis is able to focus on much lower mass YSOs, and thus bypass some of the possible biases due to sensitivity limits in far-infrared observations. The preliminary results find a star formation efficiency per free-fall time that is lower than found by \cite[Ochsendorf et al. (2017a)]{Ochsendorf:2017a}, but still in line with comparable Galactic studies such as \cite[Evans et al. (2014)]{Evans:2014}. 

Clearly more research along these lines is needed to establish whether, in addition to changes in the relation between CO and H$_2$, there are actual changes in the efficiency of the star formation process in dwarf galaxies. We discussed above expected differences in gas cooling, which may impact cloud fragmentation, when the metallicity decreases, but it may be that these differences become important at metallicities lower than those currently probed. What seems clear is that most of the molecular gas in these systems exists at extinctions that are much smaller than those in higher metallicity, more dust rich environments. Whether this matters or not is a subject of debate, as it appears that while molecules are not needed to cool a collapsing gas cloud, they are good tracers of gas capable of forming stars (\cite[Glover \& Clark 2012b]{Glover:2012b}). There exist some chemical clues hinting that star formation takes place at lower extinctions in the SMC, although it is unclear whether this results in other, deeper changes. \cite[Oliveira et al. (2013)]{Oliveira:2013} find that CO$_2$ ice features in the mid-infrared spectra of YSOs are much rarer in the SMC than in the Milky Way or the LMC, requiring a high H$_2$O ice column density threshold to be observed, with no CO ice detected in SMC sources. This suggests that the more volatile CO$_2$ and CO ices have a hard time forming (or surviving) in disks and envelopes surrounding protostars at low metallicities, at least in part because of the diminished dust shielding.

\section{The Importance of Environment}

In closing, it is important to keep in mind that besides the physico-chemical changes in the process of star formation brought about by low metallicity and low dust-to-gas ratios, there are other considerations that are particularly important for dwarf galaxies: for example, galaxies of low mass are very sensitive to their environment. Indeed, as dwarf galaxies close to the Milky Way are easily stripped from their gas through tidal and drag forces, they are also very easily perturbed by interactions with other low mass objects, which can be small galaxies or even gas clouds. A statistical relation between star formation activity and nearby companions was reported by \cite{Taylor:1997}. More recently, \cite{Stierwalt:2015} report a significant enhancement in the fraction of starbursts among close dwarf galaxy pairs versus isolated galaxies. Therefore there is mounting evidence that interactions are important in driving the ``high'' (starbursting) mode of star formation in many dwarfs, particularly that associated with the formation of massive clusters.

But past collisions, or even interactions with gas clouds, also have important and lasting effects. Deep HI observations of the dwarf starburst NGC\,1569, for example, show an HI cloud and a bridge that appear to be the remnants of a collision 1 Gyr ago that led to the starburst, which hosts a couple of super star clusters (\cite[Johnson et al. 2012]{Johnson:2012}). \cite{Lopez-Sanchez:2012} find a plume of HI in the dwarf starburst NGC\,5253 that appears to be infalling into the galaxy, likely the product of a past interaction with M\,83. \cite{Ashley:2014} find evidence for the Local Group starbursting dwarf IC\,10 either being in the latter stages of merging, or alternatively experiencing infall from a nearby HI cloud (see also \cite[Wilcots \& Miller 1998]{Wilcots:1998}). Finally, \cite{Ashley:2017} report evidence for interactions, mergers, or infall triggering several blue compact dwarfs.

Back to the Magellanic Clouds, recent GAIA measurements highlight the presence of several stellar features such as extended halos and arms that document the interaction between the Large and Small Magellanic Clouds, as well as of the Magellanic system with our galaxy (\cite[Belokurov \& Erkal 2018]{Belokurov:2018}; see also \cite[Besla et al. 2016]{Besla:2016}). The Magellanic Stream, perhaps the most spectacular feature in the HI sky, is also a result of the ongoing interaction. But does the interaction play a role in the star formation properties of the Magellanic Clouds? \cite{Besla:2012} suggest that 30 Doradus could be related to the direct collision of the SMC on the LMC. Similarly, \cite{Fukui:2017} suggests that 30 Doradus and its central massive cluster R136 are the result of colliding gas flows driven by the interaction. Perhaps even more interestingly, \cite{Ochsendorf:2017b} points out that at the other end of the LMC stellar bar, the N\,79 is ramping up its star formation and displaying the conditions to become a new supergiant HII region in the near future.  

\section{Conclusions}

The message is that, despite the advances in the last couple of decades in understanding star formation in dwarf galaxies, there remains a lot to be done to put some of the ideas on firm ground. The good news is that new facilities and those in the horizon are likely to make a substantial difference in our ability to measure some of the key parameters. The increasingly better capabilities of the Stratospheric Observatory for Far Infrared Astronomy (SOFIA), and if funded the Space Infrared Telescope for Cosmology and Astrophysics (SPICA) and the Origins Space Telescope (OST) will play a fundamental role in understanding the energetics of the ISM in the Magellanic system and beyond. When launched, the James Web Space Telescope (JWST) will allow us to carry detailed studies on the YSOs of the Local Group. Together with the Atacama Large Millimeter-submillimeter Array (ALMA) which is already making possible many observations of molecules in dwarf galaxies, future HI observatories such as the Square Kilometer Array (SKA) and the Next Generation Very Large Array (NGVLA) together with their pathfinders will allow us to explore the cold interstellar phases of the gas in these systems. The present is bright, but the future is even brighter.


\end{document}